\documentclass[a4paper,11pt]{article}
\pdfoutput=1 

\usepackage{jinstpub} 

\title{\boldmath A ROOT Based Event Display Software for JUNO}

\author[a]{Zhengyun You,}
\author[a]{Kaijie Li,}
\author[a,1]{Yumei Zhang,\note{Corresponding author.}}
\author[a]{Jiang Zhu,}
\author[b]{Tao Lin,}
\author[b]{Weidong Li}

\affiliation[a]{Sun Yat-Sen University, Guangzhou 510275, China}
\affiliation[b]{Institute of High Energy Physics, Chinese Academy of Sciences, Beijing 100049, China}

\emailAdd{zhangym26@mail.sysu.edu.cn}

\abstract{An event display software SERENA has been designed for the Jiangmen Underground Neutrino Observatory (JUNO).
The software has been developed in the JUNO offline software system and is based on the ROOT display package EVE. 
It provides an essential tool to display detector and event data for better understanding of the processes in the detectors. 
The software has been widely used in JUNO detector optimization, simulation, reconstruction and physics study.}


\begin{document}
\maketitle
\flushbottom

\section{Introduction}
\label{sec:intro}

The Jiangmen Underground Neutrino Observatory (JUNO)\cite{lab1}, a multiple purpose neutrino experiment, is currently under construction in Guangdong, China.
By precisely measuring the energy spectrum of reactor anti-neutrinos at a medium baseline of 53~km, JUNO is designed to determine the neutrino mass hierarchy.
The JUNO detector is also capable of measuring neutrinos from other different sources, such as solar neutrinos, atmospheric neutrinos, 
geoneutrinos and supernova burst neutrinos, as well as study the new physics beyond the standard model\cite{lab2}.

Event displays are valuable tools in all phases of any high-energy physics experiment.
Detector geometry and event data visualization play important roles in detector design, simulation and reconstruction algorithm tuning.
They also help problem diagnosis with online monitoring during data taking and understanding the physics in data analysis.

An event display software named SERENA (Software for Event display with Root Eve in Neutrino Analysis) has been developed for the JUNO experiment.
The motivation is to help optimizing the detector design such as photomultipliers (PMT) arrangement, understanding the simulation data, and tuning the reconstruction algorithm for complicated muons and other backgrounds with JUNO specific data model. 
The software is fully integrated into the JUNO offline framework\cite{lab3} and is based on the ROOT\cite{lab4} display package Event Visualization Environment (EVE)\cite{lab5, lab6} to implement the functions for detector and event data visualization. It also provides the functions of 3D and 2D projection views, animations, data association and interactive display for users to better understand the simulation and real data events.

In the following sections, we will first describe the design and architecture of the software, and then we will be focused on how to implement the visualization and control of the detectors, hits and their associations. Finally, the features and performance of the software will be discussed. 

\section{Design}

The JUNO offline software uses SNiPER (Software for Non-collider Physics Experiments)\cite{lab7} as its framework.
SNiPER is a light-weight, flexible framework with dependencies on external software packages such as GEANT4\cite{lab8} and ROOT\cite{lab4}.
It is designed to host all offline tasks in the experiment. 

\begin{figure}[htbp]
\centering
\includegraphics[width=12cm]{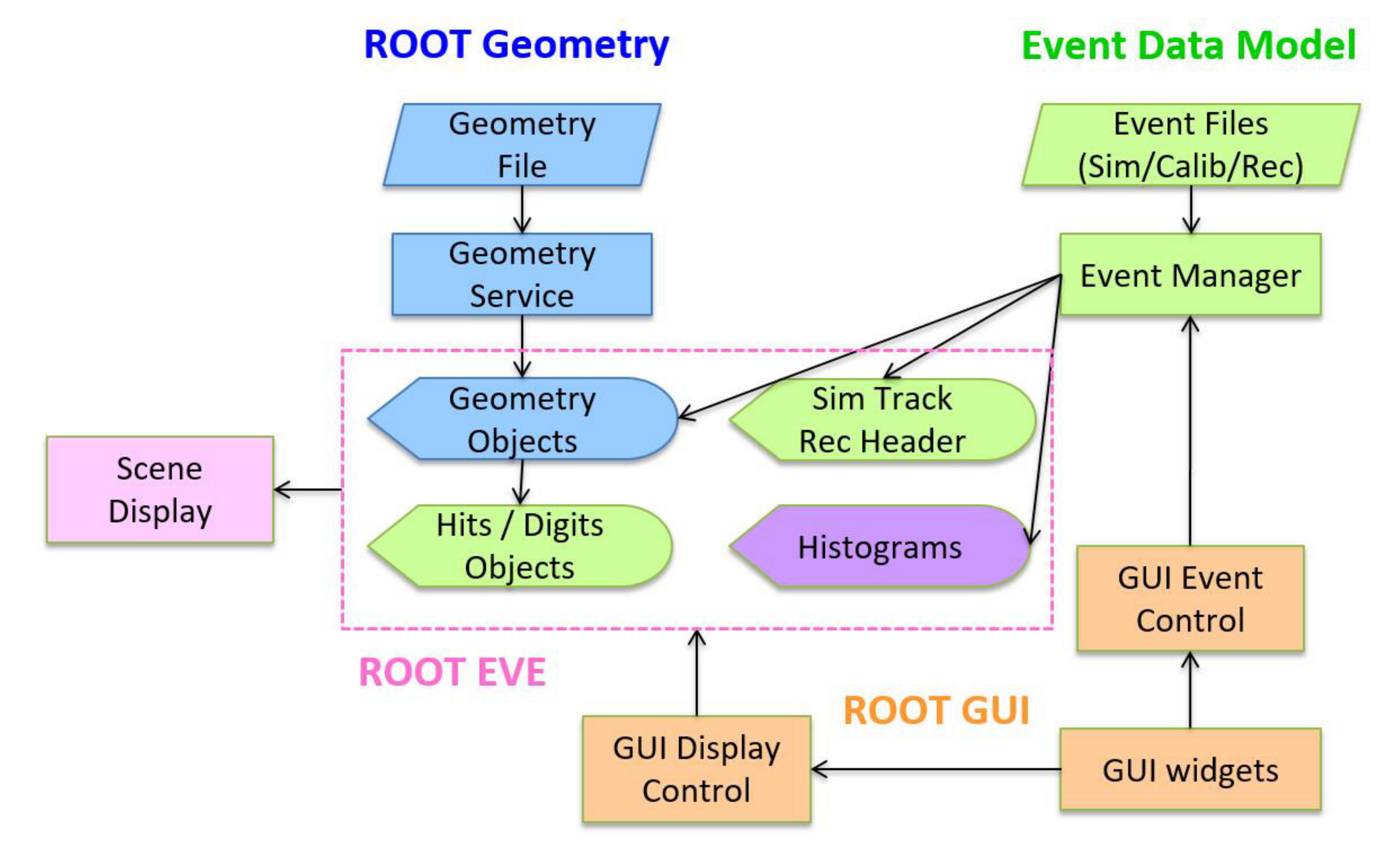}
\caption{\label{fig1} Architecture and data flow of the event display SERENA. The detector geometry is initialized as ROOT geometry objects. Event manager reads in different formats of data to get the status of detector units, particle and reconstruction information in each event. The control between users and the software is implemented with ROOT GUI package. ROOT EVE package is used to combine everything and form the scene being displayed.}
\end{figure}

SERENA is also developed in the SNiPER framework so that it can make full use of the advantages that the framework provides. The architecture and data flow in SERENA is shown in figure~\ref{fig1}. 

In the design of an event display software, three basic elements are necessary, detector geometry, event data, visualization and control of them through GUI (Graphical User Interface). 
For applications such as simulation, calibration, reconstruction and event display, SNiPER provides a common Event Data Model\cite{lab9} for data I/O and a consistent Detector Description for all applications where geometry service\cite{lab10, lab11} is used.

\subsection{Detector geometry}

\begin{figure}[htbp]
\centering
\includegraphics[width=10cm]{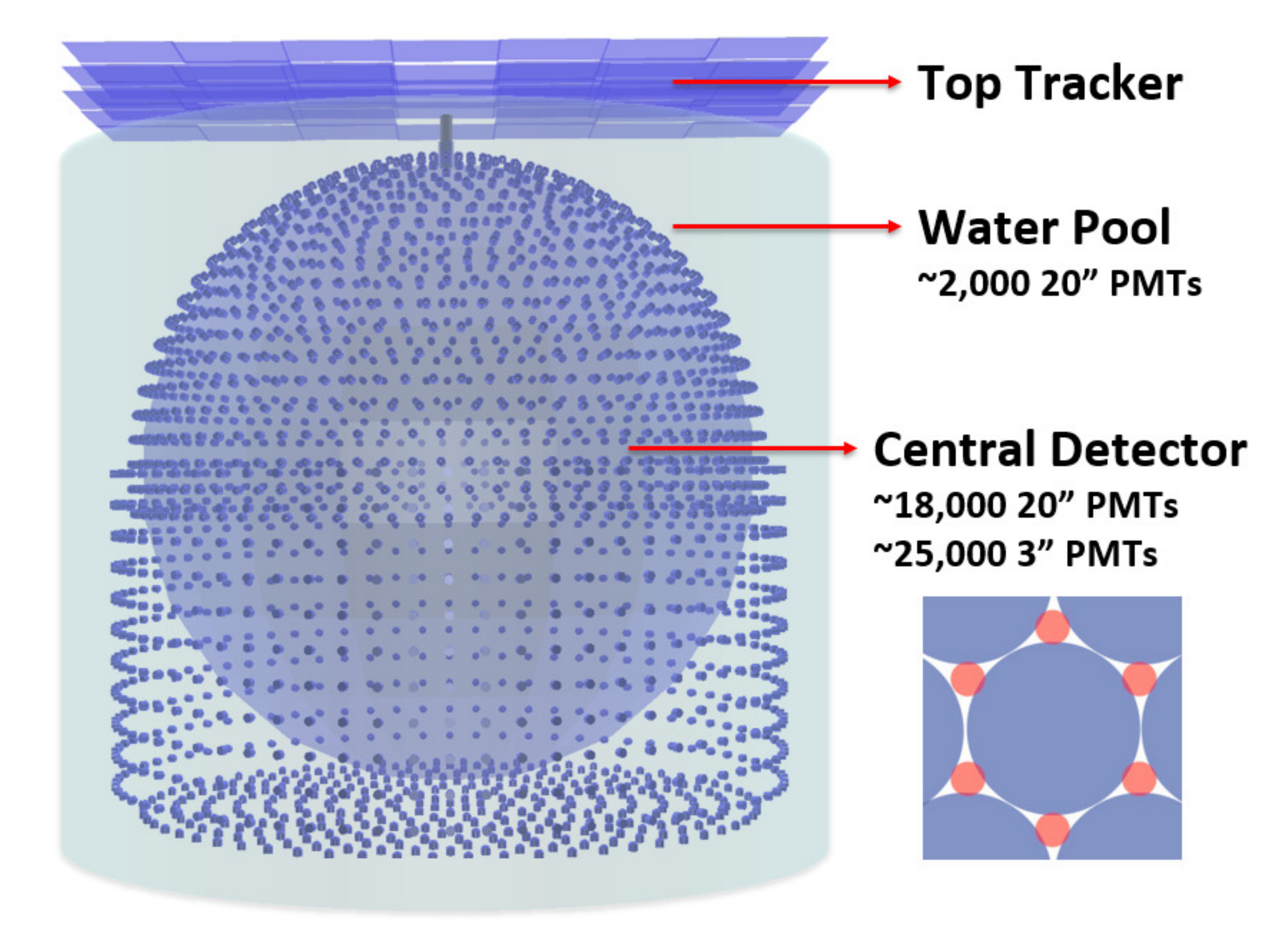}
\caption{\label{fig2} Geometry of the JUNO detector, which is made up of three sub-detectors. The Central Detector filled with liquid scintillator and surrounded with PMTs, is placed in the middle of Water Pool. On top of the Central Detector and Water Pool is the Top Tracker. The arrangement of 20-inch PMTs (large blue circles) and 3-inch PMTs (small orange circles) is shown on the bottom right corner of the figure.}
\end{figure}

The JUNO detector consists of three sub-detectors: Central Detector (CD), Water Pool (WP) and Top Tracker (TT).
The Central Detector is an acrylic sphere with radius of 17.8~m, filled with about 20~kt liquid scintillator to detect neutrinos. The surface of CD is surrounded by about 18,000 20-inch photomultiplier tubes (PMT) and about 25,000 3-inch PMTs.
The CD is placed in a Water Pool to shield from radioactive background. About 2,000 20-inch PMTs will be installed in the Water Pool to detect Cherenkov lights from backgrounds. 
On top of the WP, a Top Tracker made of plastics scintillators will be built to help muon tracks identification.
More information about JUNO detector can be found in Ref.\cite{lab1}.
Figure~\ref{fig2} shows a schematic 3D view of the JUNO detector. 

The detector geometry information in SERENA is obtained from the JUNO geometry service, which is based on Geometry Description Markup Language (GDML)\cite{lab12} and ROOT geometry package to provide a consistent detector description for different applications.
The geometry service provides a lot of convenient tools to describe the detector, such as the position, direction and boundary of every PMT, the relationship between neighboring PMTs, and the transformation between the local coordinate of every PMT itself and the world coordinate of the JUNO system. 
In SERENA, each detector unit, such as a PMT or a scintillator, is constructed with a ROOT EVE geometry object and is initialized with its own visualization attributes for later display control.  

\subsection{Event data model}

The JUNO offline software provides various event data formats at different stages, such as GenEvent from event generator, SimEvent from detector simulation, CalibEvent from calibration and RecEvent from event reconstruction.
The data are all defined by the JUNO Event Data Model (EDM) \cite{lab9} as ROOT-based persistent data objects and flow between different stages and applications.

In SERENA, the event display reads in event data files through EDM interfaces and first determines which kinds of events they are.
Different kinds of data that belong to the same event can be read in simultaneously. The event display allows reading the SimEvt, CalibEvt and RecEvt at the same time.
Every event data objects will be initialized with corresponding ROOT EVE objects,  such as the Monte Carlo truth information in simulation and event reconstruction results. 

If event data objects are related to detector units, such as simulation hits or calibration digits, instead of creating new objects, they are associated with the existing ROOT EVE geometry objects through detector identifiers.
In each individual event, the visualization attributes, such as color and visibility of the detector geometry objects, will change accordingly. 

\subsection{Graphical User Interface (GUI)}

\begin{figure}[htbp]
\centering 
\includegraphics[width=14cm]{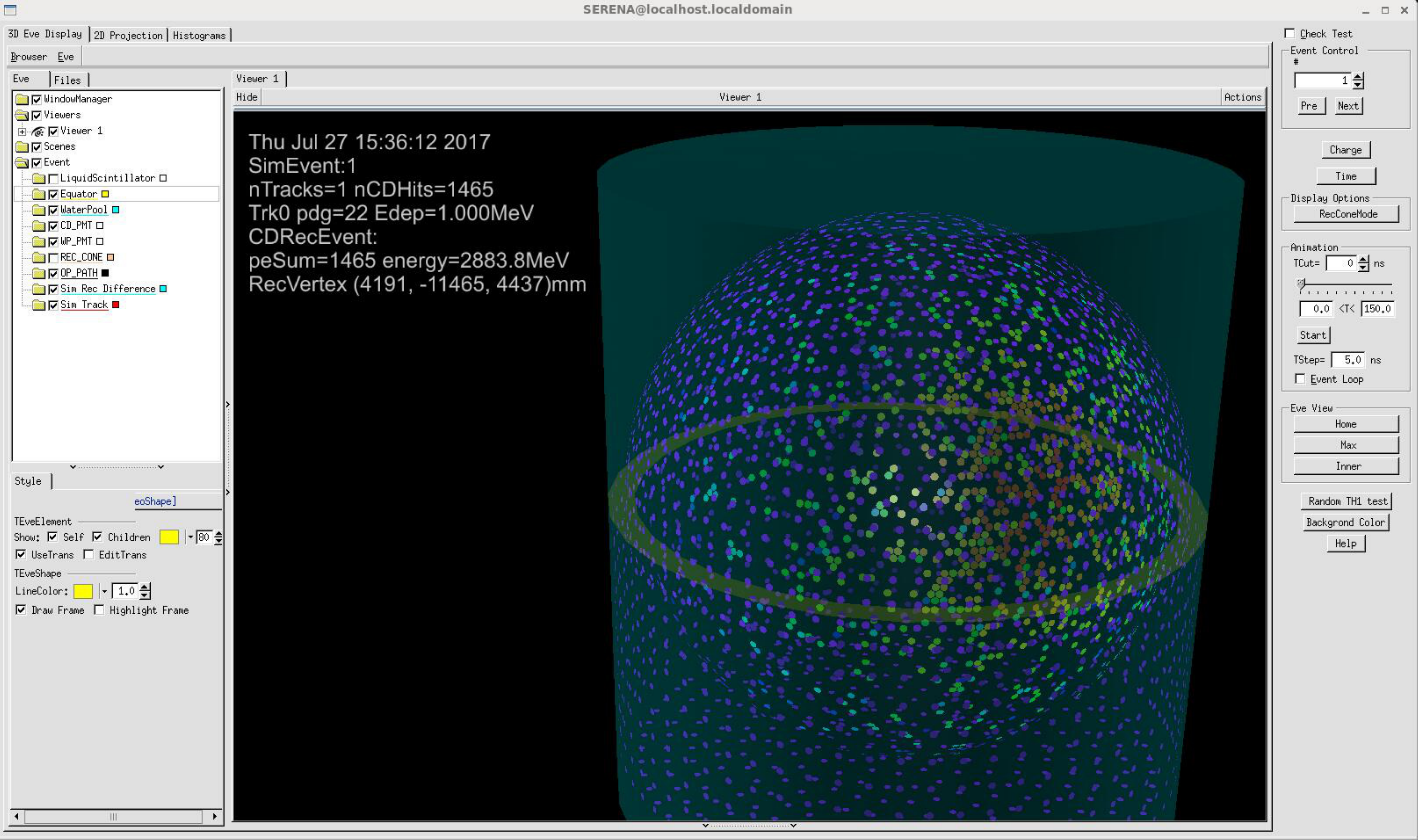}
\caption{\label{fig3} The Graphical User Interface of SERENA.}
\end{figure}

The Graphical User Interface in event display is designed for users to realize the interactive control of detectors, event data and scene display.
The GUI in SERENA is implemented in combination of ROOT EVE and ROOT GUI package, as shown in figure~\ref{fig3}. 
Besides of the default functions provided by EVE, the GUI should also be able to provide easy control of the geometry, view, hits and event flow, as well as the implementation of animations.

In figure~\ref{fig3}, the center of the window is the main display frame, where the detectors, hits and tracks are plotted. 
It uses the ROOT GUI and OpenGL modules\cite{lab13} for data presentation and interaction layer.
On the top left corner of the main display frame is a text box, which shows the important summary information of the current event being displayed, such as event time, number of hits, reconstructed vertex and energy, and so on. 
On the left panel of the window, a default GUI window is implemented by ROOT EVE for file navigation, detector objects visibility switch, visualization attributes setting and user-feedback. 
One the right panel, a series of GUI widgets including buttons, sliders, check box and  text boxes are designed to realize the functions such as event navigation, animation, camera control and histogram display.
The GUI also makes it possible for developers and data analyzers to register the objects to display one by one, so as to inspect the intermediate results in complicated processes.   

\section{Visualization}

Visualization of the detector units and event elements are the essential functions of event display software. 
The main display window of GUI in ROOT-EVE is based on the OpenGL graphic engine to realize the visualization effects.
In SERENA, we have implemented several functions to display the elements as described below.

\subsection{Detectors}

The descriptions of all sub-detectors are converted from GDML and are based on ROOT geometry. It provides access to detailed description of PMTs, including materials, shapes and coordinates information.
However, visualization of such fine descriptions for all detector units are impractical in event display since there are more than 45,000 PMTs in the whole JUNO detector.
The graphical processing power requirements are beyond most end-users' computer hardware limits.
Optimizations have been applied to detector units. For example, PMTs are simplified into a set of TEveBox cone objects, which makes it possible to draw all PMTs and to keep smooth updating of the drawings in animations on most users' laptop.     
With such implementation, the visualization speed and response of the software is greatly improved.  

\begin{figure}[htbp]
\centering 
\includegraphics[width=8cm]{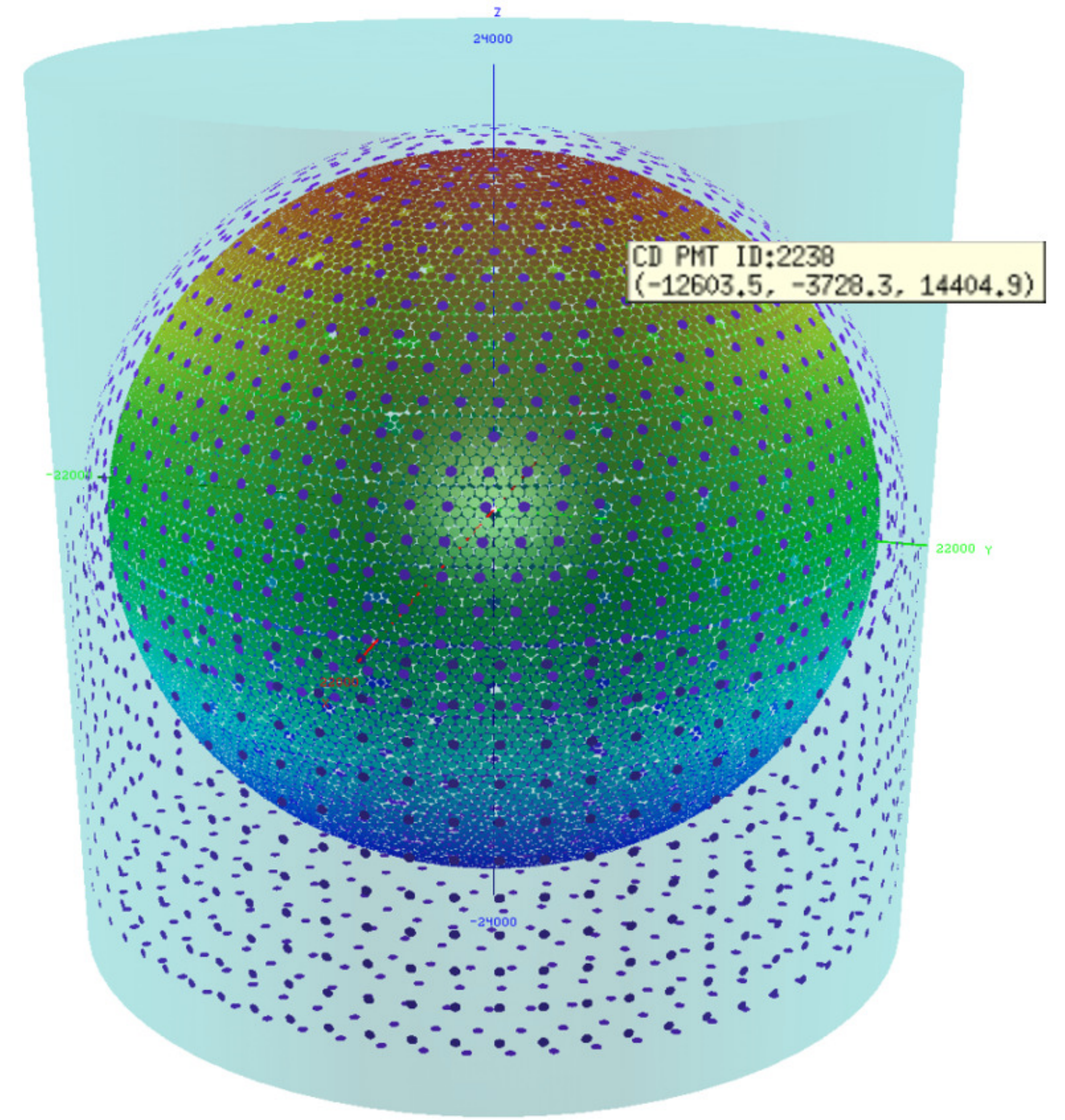}
\caption{\label{fig4} Display of the detector geometry. The light blue cylinder is the Water Pool. The purple points in the Water Pool are veto PMTs. The colorful sphere is made of PMTs in the Central Detector, with different colors representing the position of every PMT.  When the cursor is moved on top of a PMT, a text box pops up to show its information including the position and identifier number.}
\end{figure}

Figure~\ref{fig4} shows the visualization effects of the whole Central Detector and Water Pool with all 20-inch PMTs turned on.
Every detector object is selectable and pickable. For example, when the cursor is moved on top of a PMT, it is picked and a text box with its identifier and position information pops up automatically. Then the users can select this detector object to change its visualization attributes. 
The users have the flexibility to decide whether to display some detector units or not, and change the visualization attributes of some specific detector units. 

\subsection{Hits distribution}

\begin{figure}[htbp]
\centering 
\includegraphics[width=8cm]{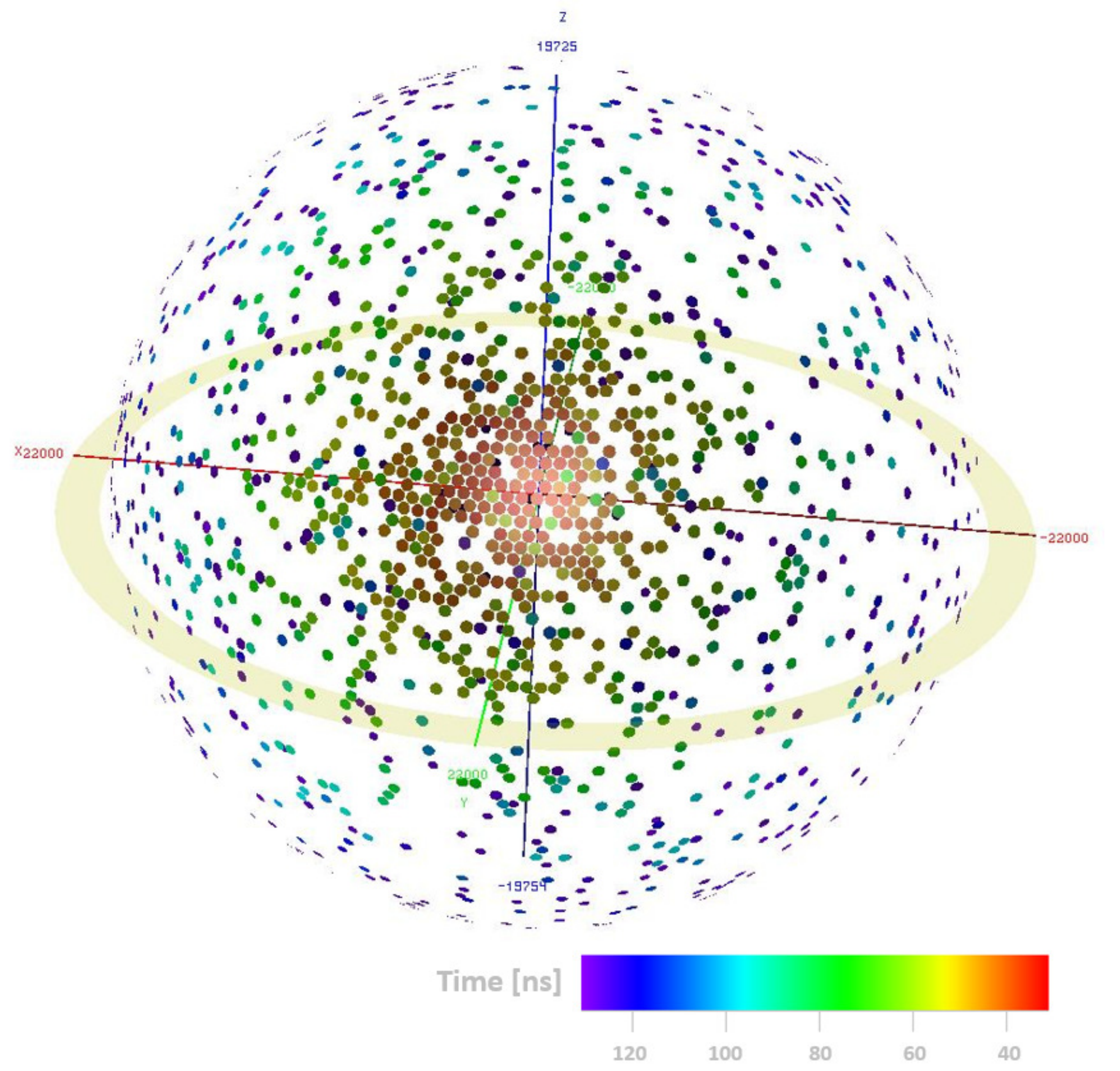}
\caption{\label{fig5} Time distribution of PMT hits in an event, represented with different colors. The red PMTs are closer to the vertex with early hit arrival time, while the purple PMTs are far from the vertex with longer traveling distance and late hit arrival time.}
\end{figure}

One of the main tasks in JUNO event display is to show the hits distributions on PMTs in each event.
SERENA provides two modes to show the distribution of hits, number of photon hits on a PMT (charge distribution) and the earliest hit time on a PMT in an event (time distribution), as shown in figure~\ref{fig5}.

The charge distribution and hit time distribution play important roles for software developers to tune the reconstruction algorithm. For example, to reconstruct the event vertex, in the charge weight center reconstruction method, the event vertex is determined by weighted center of hits charge distribution on all PMTs, while in the hits time reconstruction method, the algorithm tries to find the maximum likelihood position of event vertex by calculating the transportation time from the vertex to every PMT hits\cite{lab1}. The event display of the charge and time distribution provides an intuitive view to understand the algorithms.

\subsection{Associations}

In reconstruction algorithm tuning and physics analysis, the users usually require comparing the reconstruction results with MC truth.
SERENA can read in different kinds of data format from the same event and display them on the same scene simultaneously, so that the users can compare the associations between them. 

\begin{figure}[htbp]
\centering 
\includegraphics[width=14cm]{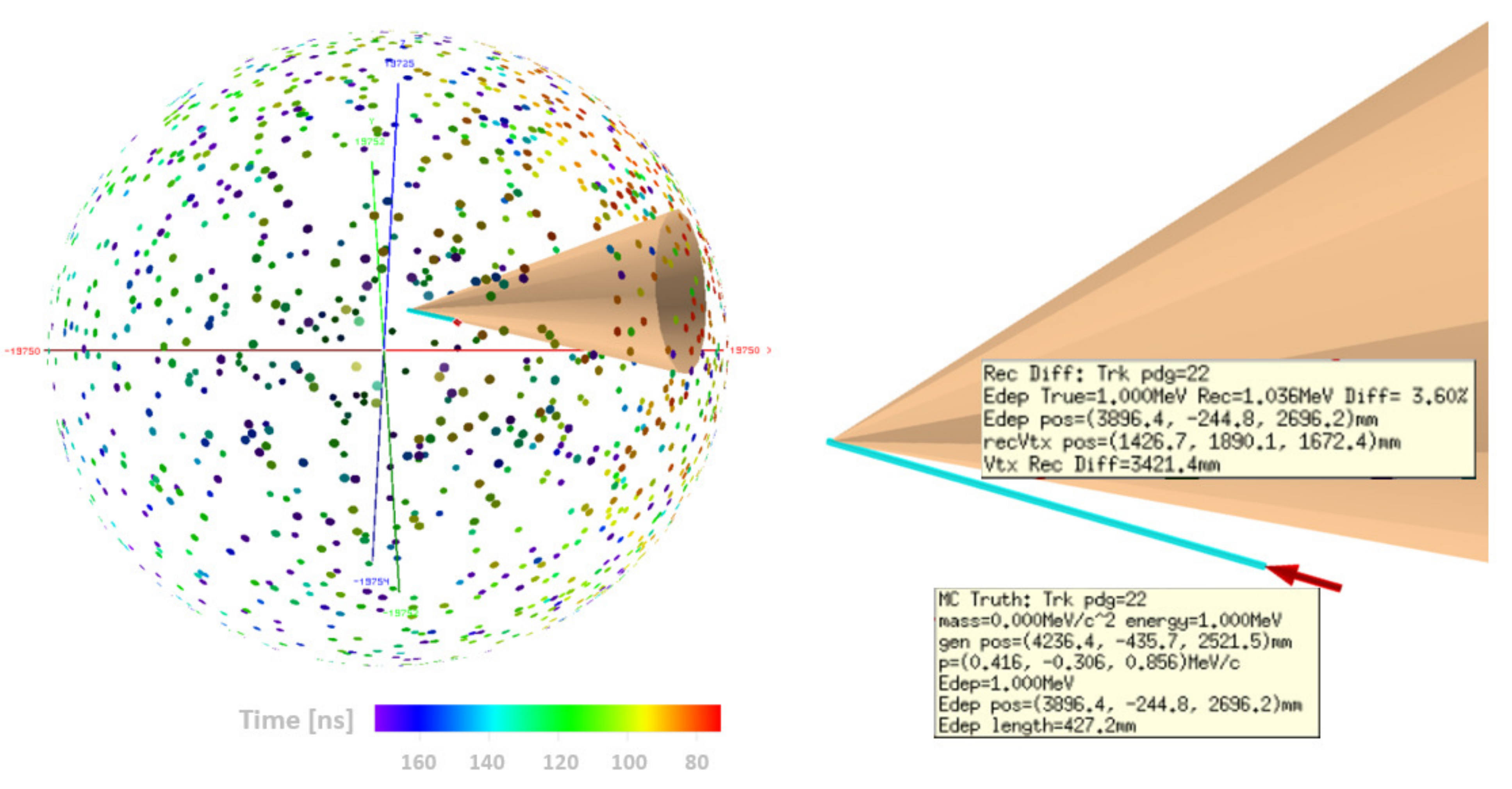}
\caption{\label{fig6} Comparison of the reconstructed vertex and the true vertex in a Monte Carlo gamma event. In the left figure, the point at the peak of the cone tells the users where the reconstructed vertex is. The right figure is the zoom-in view of the left figure near the vertex. Gamma is generated at the tail of the red arrow and deposits its energy at the head of the red arrow, which is the true vertex of the event. The bottom text box shows its MC truth information. The point at the peak of the cone is the position of reconstructed vertex. A blue line connects the head of red arrow and peak of the cone. It tells users the result of reconstruction with information shown in the top text box.}
\end{figure}

Figure~\ref{fig6} shows the difference between reconstructed vertex and true vertex in a Monte Carlo gamma event.
The point at the peak of the cone is the reconstructed event vertex from RecEvent.
The tail and head of the red arrow, as shown in the zoom-in region, represent the production vertex of gamma and its energy deposit position from Monte Carlo truth in SimEvent, respectively. 
A blue line connecting peak of the cone and red arrow head tells users the difference between reconstruction vertex and its true position.
The shorter the blue line is, the better the event vertex is reconstructed. 
For a perfect vertex reconstruction, the length of the blue line is zero.
SERENA also provides the functions to compare MC truth in different sub-detectors like PMT hits in Central Detector and those in Water Pool.

\subsection{Animations}

Besides static pictures, some modern event display software provides animation function to give the users a vivid view of the event.
Essentially, animations are a set of pictures displayed by continuously refreshing the window with time evolution.
Animations are useful tools to understand how physical events evolve from the beginning to the end, how particles are generated, propagate in the materials, interact with sensitive detectors and produce hits.

\begin{figure}[htbp]
\centering 
\includegraphics[width=14cm]{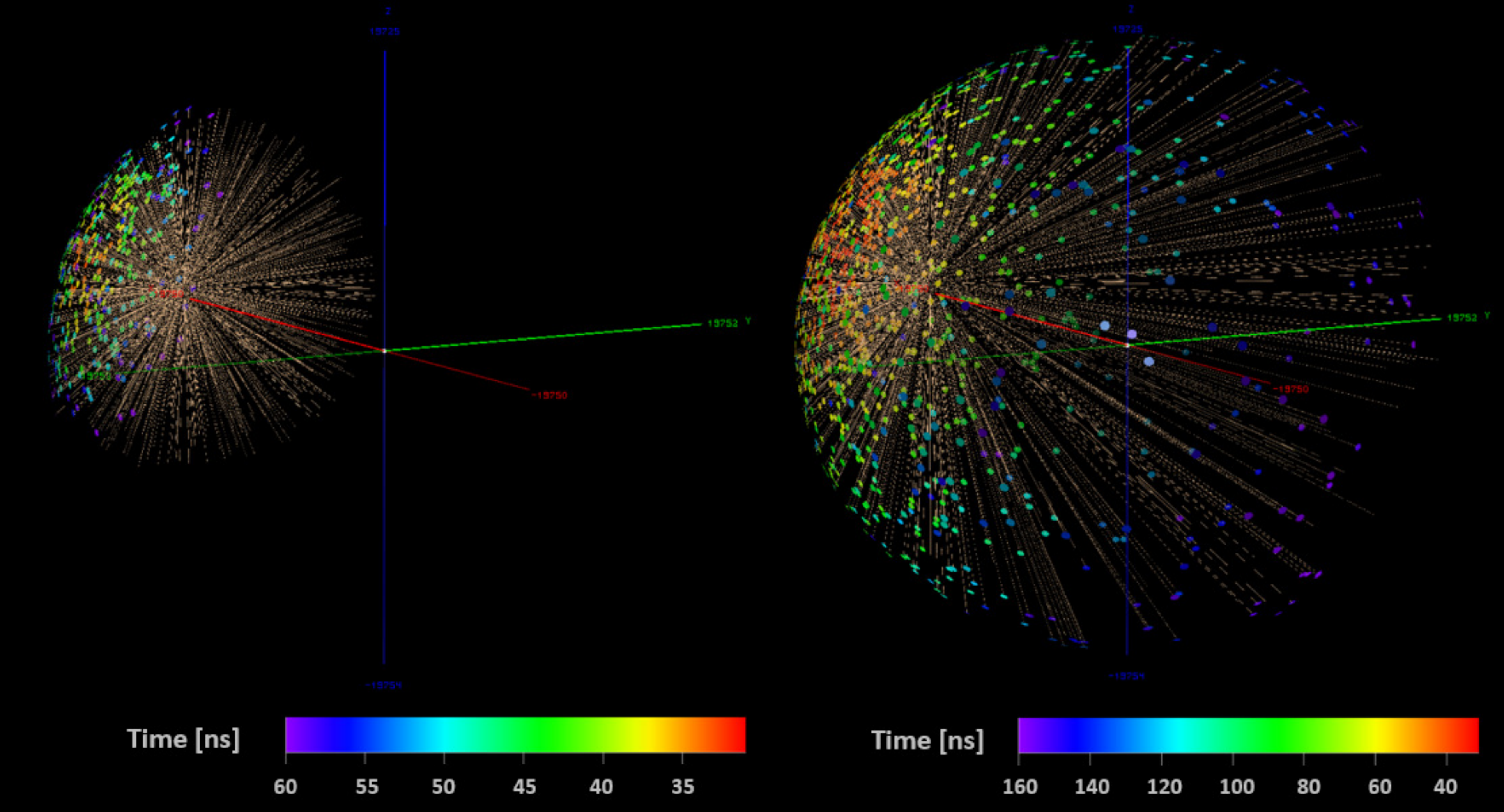}
\caption{\label{fig7}  Animation of photons propagation in the Central Detector. The two plots show the same event at different time. The left plot shows the propagation of photons at the time of 60 ns. Some photons on the left side have already reached the PMTs and fired them, while the other photons are still propagating. The right plot shows the propagation of photons at the time of 160 ns. Almost all photons have reached the PMTs.}
\end{figure}

In figure~\ref{fig7}, the propagation of photons in the liquid scintillator of the Central Detector at two intermediate moment (60 ns and 160 ns) are shown. Some photons have reached the surface of the Central Detector and produced hits on PMTs, while the other photons are still propagating. 

In SERENA, the animation function is realized by displaying the hits with different time cuts.
Users have the freedom to control the animation configurations by setting  the time window range, whether to replay or not, and the play speed by changing the time interval between every display refreshing.
In combination with the ROOT-EVE camera rotations, zooming and movements, more fancy animation effects can be realized easily.

\subsection{2D projection view}

Most of the event display with ROOT-EVE are three-dimensional (3D) views. But sometimes two-dimensional (2D) projection views are more helpful for users to understand a physical event\cite{lab14}.
In ROOT-EVE, the projections are performed automatically on extracted geometry, points, tracks and hits. Currently EVE provides r-$\varphi$ view and $\rho$-z view for default 2D projections.

\begin{figure}[htbp]
\centering 
\includegraphics[width=10cm]{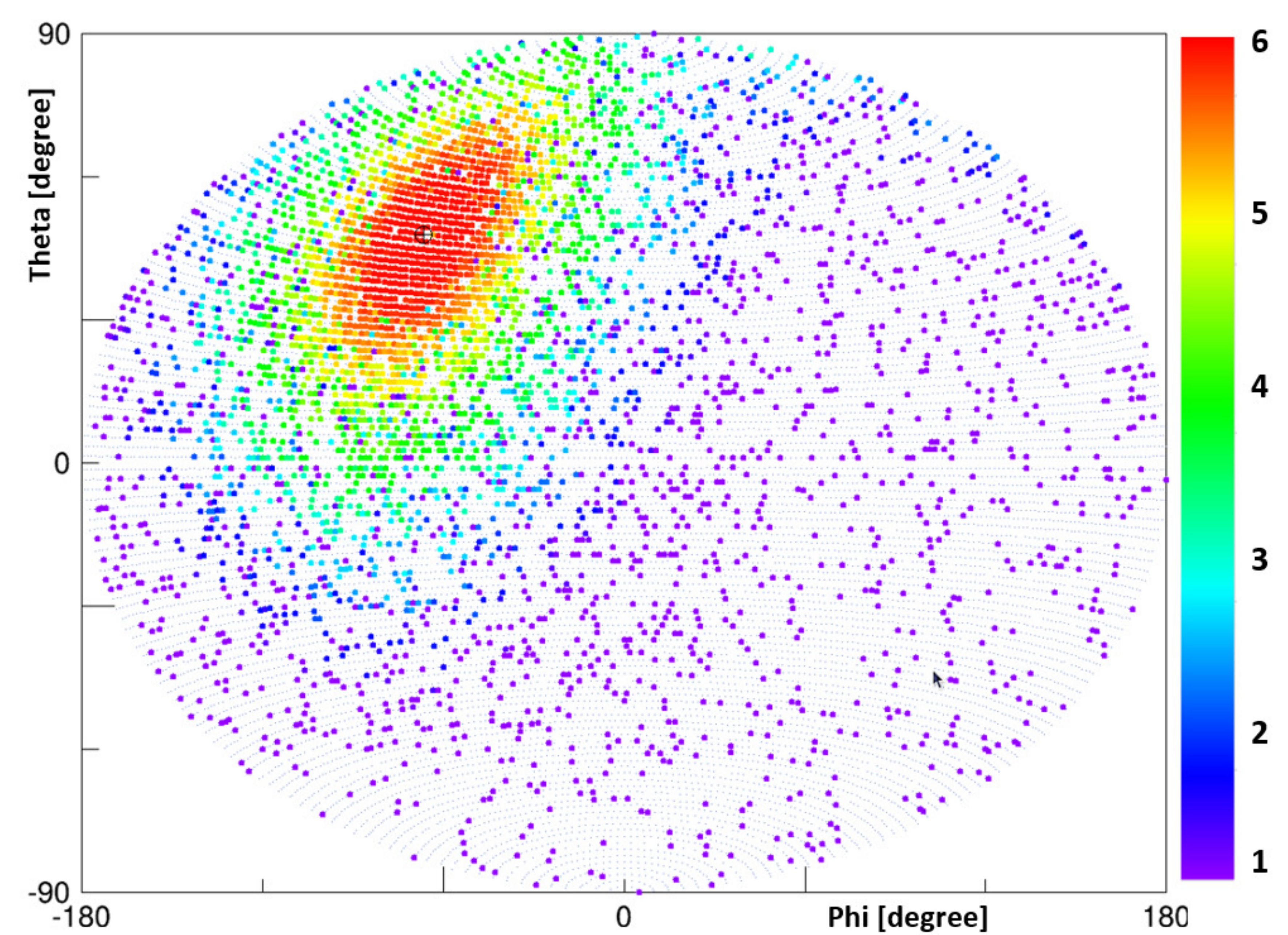}
\caption{\label{fig8}  2D projection ($\theta$-$\phi$) of the Central Detector PMT hits charge in an event. The red points are PMTs with maximum number of hits, while the purple points are PMTs with only one hit.}
\end{figure}

In the world coordinate of the JUNO detector system, the z coordinate is vertically pointing to the sky, $\theta$ and $\phi$ are the default polar angle and  azimuthal angle defined in spherical coordinate, respectively. 
Since the shape of the JUNO Central Detector is almost a symmetric sphere, the default 2D projections in EVE does not give much helpful information of the hits distribution.
Instead, it is more helpful to use the Atioff 2D projection, which projects the whole surface of a sphere onto a 2D plane plot like flattening the Earth to a world map\cite{lab15}, as shown in figure~\ref{fig8}.

In figure~\ref{fig8}, the solid circle markers with different colors represent the fired 20-inch PMTs with different number of photon hits, and the background grey points are PMTs without any hit on it.
The black cross and circle represent the projected vertex position from reconstruction output and MC truth, respectively.

\section{Features}

A few features have been implemented in SERENA to provide clear and convenient interactions between the display and users. 

1) Interactive display

The users can rotate, move and zoom in/zoom out a scene to get a best view.
Every ROOT-EVE objects are pickable and selectable (i.e. figure ~\ref{fig4}). 
Selective data display plays an important role in the design of SERENA.
The users decide which object or object collections to display by turning on/off them, so that they will not be distracted by other irrelevant event contents. 
If the users are particularly interested in some specific objects, they can change the visualization attributes of these objects to emphasize them.

2) Object filtering

Several criteria are implemented as cuts to filter objects in display.
Users can select to only display the objects that pass the regular expression of the cuts. 
For example, visualization of the PMT hits within a time window or charge range can be filtered by implementation of the time cut  (i.e. figure ~\ref{fig7}) or charge cut. 

3) Tooltip

In data analysis or algorithm tuning, the users may need to know more details about an object rather than what is displayed (i.e. figure ~\ref{fig6}).
Tooltips are implemented by attaching text information to an EVE object.
When a mouse cursor is moved on top of an EVE object, such as a hit, a track or a detector unit, a text box will pop up to display its detailed information.
For example, moving the cursor on a PMT will show its detector identifier and center position.   
By moving the cursor on top of the reconstruction cone, much more information, such as the number of hits on it, reconstruction vertex and reconstruction energy of this event will be shown.
Besides, moving the mouse cursor on top of any GUI button explains the function of this widget to help the users better understanding it.

4) Associations

The ability to display different formats of data and their associations in the same scene is one of the unique features of SERENA.
Comparisons between calibration data, reconstruction output and Monte Carlo truth are convenient with this feature being implemented (i.e. figure ~\ref{fig6}). 

5) Animations
  
 Animations help the users, especially the beginners, to understand how different signal and background events are formed. 
 It also provides a vivid view of the physics processes from the simulation data of an event (i.e. figure ~\ref{fig7}). 

\section{Performance}

For the most common Inverse Beta Decay (IBD) events in JUNO, SERENA has the ability to process events promptly and refresh the display more than 10 times per second in test running on a local computer with Intel Core i7 CPU.
The typical total memory consumption for SERENA to display IBD events is about 300 MB, with simulation, calibration and reconstruction data files opened.
When users run the software remotely through internet, the response of interactions mostly depends on the latency between the users and server.

\begin{figure}[htbp]
\centering 
\includegraphics[width=8cm]{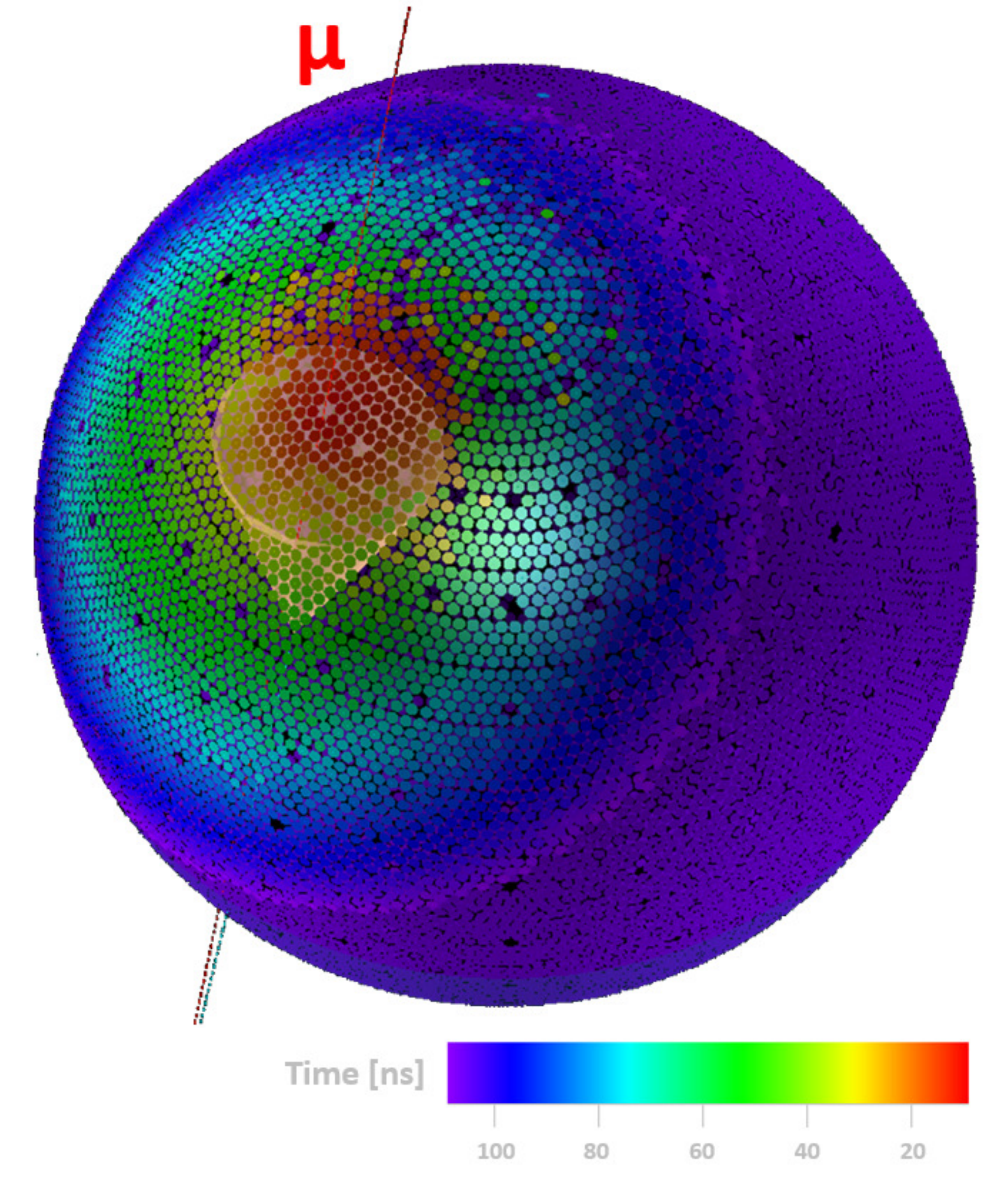}
\caption{\label{fig9}  Time distribution of PMT hits in a high energy cosmic ray muon event. A muon with momentum of 200 GeV/c penetrates the whole Central Detector from top to bottom and deposits about 1GeV energy in the liquid scintillator. Red hits arrive at PMTs early, while purple hits arrive late.}
\end{figure}

A typical neutrino IBD event produces about thousands of PMT hits. 
However, cosmic ray background muons, especially high-energy muons penetrating the whole JUNO detector, will produce millions of hits in one event and fire all PMT in a short time.
This makes it a challenge to display such events, not only in data I/O, but also in computing and graphic processing. 
For events with a huge number of photons, not every hit is traced and displayed. The hits that belong to the same PMT are clustered and are sorted to give a summary hit information of the PMT for visualization.
With algorithm optimization and data simplification, we have made it possible to draw a muon event with millions of hits in one second (figure \ref{fig9}). 

Scientific Linux is the most commonly used system where JUNO offline software runs on, SERENA has been tested and run smoothly on remote clients such as Linux, Windows and Mac systems.
The event display software is fully integrated in the JUNO offline software, which makes it convenient to connect with the online DAQ system to make an online monitoring or event display after some modifications.

\section{Conclusions}

We have developed an event display software SERENA in the JUNO offline software.
It is based on the SNiPER framework and the ROOT EVE environment.
In the past 4 years of JUNO design and offline software development, SERENA has proven to be an essential tool in detector optimization, simulation, reconstruction and aiding physics analysis.
It also provides a good material for outreach of the JUNO experiment and to present the physics of JUNO to the public. 
A few versions of the event display have been successfully released and more functions are expected to be added in the future updates.

\acknowledgments

The authors would like to acknowledge the supports provided by National Natural Science Foundation of China (11405279, 11675275), the Strategic Priority Research Program of Chinese Academy of Sciences (XDA10010900) and the Recruitment Program of Global Experts in China.
The authors would also like to thank members of the JUNO offline software group for their valuable discussions and suggestions, and thank the JUNO simulation and physics analysis collaborators for their feedbacks.


\end{document}